# 4MOST Consortium Survey 3: Milky Way Disc and Bulge Low-Resolution Survey (4MIDABLE-LR)

Cristina Chiappini[1]
Ivan Minchev[1]
Else Starkenburg[1]
Friedrich Anders[2]
Nicola Gentile Fusillo[3]
Ortwin Gerhard[4]
Guillaume Guiglion[1]
Arman Khalatyan[1]
Georges Kordopatis[5]
Bertrand Lemasle[6]
Gal Matijevic[1]
Anna Barbara de Andrade Queiroz[1]
Axel Schwope[1]
Matthias Steinmetz[1]
Jesper Storm[1]
Gregor Traven[7]
Pier-Emmanuel Tremblay[3]
Marica Valentini[1]
Rene Andrae[8]
Anke Arentsen[1]
Martin Asplund[9]
Thomas Bensby[7]
Maria Bergemann[8]
Luca Casagrande[9]
Ross Church[7]
Gabriele Cescutti[10]
Sofia Feltzing[7]
Morgan Fouesneau[8]
Eva K. Grebel[6]
Mikhail Kovalev[8]
Paul McMillan[7]
Giacomo Monari[1]
Jan Rybizki[8]
Nils Ryde[7]
Hans-Walter Rix[8]
Nicholas Walton[11]
Maosheng Xiang[8]
Daniel Zucker[12]
and the 4MIDABLE-LR Team

[1] Leibniz-Institut für Astrophysik Potsdam (AIP), Germany
[2] Departament de Física Quàntica i Astrofísica, Universitat de Barcelona, Spain
[3] Department of Physics, University of Warwick, UK
[4] Max-Planck-Institut für extraterrestrische Physik, Garching, Germany
[5] Observatoire de la Côte d'Azur, Nice, France
[6] Zentrum für Astronomie der Universität Heidelberg / Astronomisches Rechen-Institut, Germany
[7] Lund Observatory, Lund University, Sweden
[8] Max-Planck-Institut für Astronomie, Heidelberg, Germany
[9] Research School of Astronomy & Astrophysics, Australian National University, Canberra, Australia
[10] Osservatorio Astronomico di Trieste, INAF, Italy
[11] Institute of Astronomy, University of Cambridge, UK
[12] Department of Physics and Astronomy, Macquarie University, Sydney, Australia

The mechanisms of the formation and evolution of the Milky Way are encoded in the orbits, chemistry and ages of its stars. With the 4MOST MIlky way Disk And BuLgE Low-Resolution Survey (4MIDABLE-LR) we aim to study kinematic and chemical substructures in the Milky Way disc and bulge region with samples of unprecedented size out to larger distances and greater precision than conceivable with Gaia alone or any other ongoing or planned survey. Gaia gives us the unique opportunity for target selection based almost entirely on parallax and magnitude range, hence increasing the efficiency in sampling larger Milky Way volumes with well-defined and effective selection functions.

## Scientific context

Observations of star-forming regions suggest that stars are born in groups that possess a high degree of chemical homogeneity. Large-scale galactic dynamical processes (spiral arms, the central bar, mergers) affect these stellar aggregates kinematically but not chemically. Comparison between dynamical models of the Galaxy and the observed velocity field for a densely populated homogeneous disc area will allow us, for the first time, to unambiguously quantify the structure of the present-day bar, spiral arms, and asymmetries across the disc mid-plane. By combining this information with accurate elemental abundances and ages, we will reconstruct the enrichment of the Galaxy as a function of Galactic radius and time. With its high multiplex power, 4MOST will allow us to view the Milky Way as a whole.

With the Low-Resolution Spectrograph[a] (LRS) of 4MOST we can study the origin and evolution of essentially all the dominant stellar components of the Milky Way: the chemically and structurally defined thin and thick discs, and the bulge. The area covered by the 4MIDABLE-LR, as well as by the 4MOST MIlky Way Disc And BuLgE High-Resolution survey for brighter stars (4MIDABLE-HR) will be large enough to also enable a comprehensive study of the disc/bulge, disc/halo, and bulge/halo interfaces for the first time — the latter two in collaboration with the 4MOST Consortium Milky Way Halo LR Survey.

## Specific scientific goals

Our main goals are:
– To better understand the current Milky Way disc structure and dynamics (bar, spiral arms, vertical structure, stellar radial migration, merger history).
– To study the chrono-chemo-dynamics of the disc, which, when combined with the above will allow us to recover the disc evolutionary history.
– To better understand the formation of the Milky Way bulge/bar using both chemical and kinematical information.
– To study the inner disc/bulge and disc/halo interfaces by covering a large area and ensuring high-quality chemical and kinematical information.

These goals could be summarised as one main goal — to provide a detailed chrono-chemo-kinematical extended map of our Galaxy and the largest Gaia follow-up down to $G = 19$ magnitudes (Vega). The complex nature of the disc components (for example, large target densities and highly structured extinction distribution in the Milky Way bulge and disc area), as well as the nature of the open questions addressed above, prompted us to develop a survey strategy with five main sub-surveys that are tailored to answer the main science questions, while taking full advantage of the Gaia data.

## The main sub-surveys are:

1. **Extended Solar Neighbourhood (ESN)** — The aim of this sub-survey is to study in detail the chemistry and velocity substructure in a disc volume for which Gaia provides the most precise parallaxes ($d < 2$ kpc). Because we select stars that are fainter than those that have been targeted for spectros-copy by Gaia, we target this region with many more and different tracers, including sub-giant stars (~ 20%) which are suitable for accurate age determination. Additionally, we add elemental abundances not obtainable from Gaia spectra, such as the *n*-capture elements Ba and Sr. Constraining the dynamical and chemical state in this nearby region will drastically improve our understanding of the influence that the bar, spiral arms, and external perturbations (for example, from the Sagittarius dwarf spheroidal galaxy) have on the disc dynamics and stellar radial migration (Minchev, Chiappini & Martig, 2013). This is our brightest sub-survey and the most densely sampled region.

2. Dynamical disc (Dyn) — This sub-survey targets a dense disc sample to detect velocity resonance structures, stellar streaming, and disc ringing in order to better understand the current structure and dynamics of the Milky Way disc (Minchev, 2016). Comparing the velocity field in different parts of the Galaxy will allow us to unambiguously determine the physical mechanisms perturbing the Milky Way in a way that is not possible with smaller or patchy survey volumes. It is therefore crucial that we observe a dense sample over multiple (four to five) disc scale-lengths, leveraging the unique capabilities of 4MOST, and produce an extended map to facilitate comparisons between the well-studied Milky Way and galaxies in general.

3. Faint dynamical disc (Fdyn) — This sub-survey has an overall goal similar to that of the Dynamical disc sub-survey, but with the explicit aim of studying the stellar disc to its edge, and also at the other side of the Galaxy (out to $d \sim 30$–40 kpc) providing mainly radial velocity information rather than elemental abundances for these faint targets. This will be a unique capability of 4MOST, even in the 2020s.

4. Chemodynamical disc (Chem) — Here we target a sparser sample as a subset of the Dyn sub-survey described above, to study the chemodynamical structure of the Milky Way disc in order to recover the evolutionary history of the disc. Higher signal-to-noise (S/N) observations enable measurement of elemental abundances for many stars for iron, carbon, several alpha elements and iron-peak elements, lithium, sodium, and the n-capture elements Ba and Sr. Our data will allow us to constrain radial and vertical chemical gradients, as well as velocity dispersions as a function of stellar age when combined with sub-samples of stars for which ages can be reliably determined (from asteroseismic measurements, or otherwise). The large sample of stellar spectra will also allow identification of stars that were born in now-dissolved star clusters and associations via chemical labelling and kinematics, when the cluster chemistry and/or kinematics are sufficiently different from the field population. Moreover, it may lead to the serendipitous discovery of extra-tidal stars around surviving clusters, imposing constraints on dynamical mass loss.

5. Bulge/Inner Galaxy (BIG) — To better understand the formation of the Milky Way bulge/bar region and to study the inner disc/bulge/halo interfaces, this sub-survey will map the inner 4–5 kpc of the Galaxy, encompassing the full bar length and peanut thickness, which is still poorly covered by other spectroscopic surveys (Barbuy et al., 2018). This will certainly be a legacy survey, as we aim to provide the largest coverage of the inner disc/bulge volume, focusing on the interplay of all different Galactic components in this region. BIG will fully complement the surveys with the Multi-Object Optical and Near-infrared Spectrograph (MOONS), the Apache Point Observatory Galaxy Evolution Experiment (APOGEE), Sloan Digital Sky Survey V, and the 4MOST MIlky Way Disc And BuLgE High-Resolution (4MIDABLE-HR) survey. We will be taking advantage of complementary photometry from the ESO Vista Variables in the Via Lactea Surveys (VVV and VVVx) in the $JHK_s$ bands and of the Blanco DECam (Dark Energy Camera) Bulge Survey (BDBS) which covers 200 square degrees of the southern bulge in the *ugrizY* bands and will be publically available by the end of 2019.

Whilst most of our survey is carried out under bright sky conditions, some dark and grey time will be used for this sub-survey. Moreover, in a joint effort with S4, we plan a 4MOST southern bulge deep-field sub-survey with a grid of pointings in the region $-8 < l < 8$ degrees and $-10 < b < -4$ degrees to be observed for 8 hours each. This will allow us to extend the LR coverage down to $G$ = 18.5–19.0 magnitudes in these fields, hence complementing our main survey in the $16 < G < 17$ magnitude range.

In parallel with our main sub-surveys, we will have the following six sub-surveys targeting specific classes of stars. These can all be pursued simultaneously and efficiently thanks to the multiplexing capabilities of 4MOST and the fact that they consist of targets sparsely distributed over the footprint of the main surveys.

6. Very metal-poor stars — These are important tracers of the early evolution of the Galaxy. We will select these targets in the disc and inner Galaxy from additional photometric surveys (Sky-Mapper and the Pristine survey), and from APOGEE. For the latter targets spectroscopy is available, but for this type of target the 4MOST optical wavelength coverage rather than the (near-) infrared APOGEE wavelength region is necessary to obtain accurate elemental abundance information and study their chemical pattern. Our very metal-poor sub-survey complements the 4MOST Consortium Milky Way Halo High-Resolution Survey, which is also pre-selecting metal-poor targets, as we are probing fainter targets with the LRS and are looking in the inner Galaxy and in the disc footprint. Some overlap is desired for calibration purposes.

7. White dwarfs — This sub-survey will complement the Gaia mission and other Galactic sub-surveys enabling the use of white dwarfs as tracers of the star formation history in the disc and halo, which can help to disentangle different scenarios of stellar radial migration. Additionally, these observations will provide key constraints on the nature of SN Ia progenitors and the evolution of planetary systems. The goal is to determine atmospheric compositions, radial velocities, and precise spectroscopic temperatures and surface gravities, which, in turn, will allow us to determine accurate stellar masses and ages for all targeted white dwarfs.

8. **Compact X-ray emitting binaries** — This sub-survey addresses the high-energy output of the Milky Way and the evolution of close binary stars. We aim to disentangle accreting and non-accreting sources, and to discriminate between magnetic and non-magnetic binaries, both on the basis of their emission line spectrum in the active state (for example, Schwope, 2018). Input sources will be drawn from the eROSITA point-source catalogue.

9. **Cepheids** — Classical Cepheids have well-determined ages (~ 20–250 Myr). They are intrinsically bright, thus allowing studies of the recent Milky Way evolution to large Galactic radii. Type II Cepheids are old (> 10 Gyr) and trace the chemical thick disc and its interface with the bulge. Targets will be selected from the ESA Gaia data releases, the Optical Gravitational Lensing Experiment (OGLE), and VVVX. The goal is to obtain homogeneous radial velocities and metallicities/elemental abundances for these unique tracers down to fainter magnitudes than those achieved with 4MIDABLE-HR.

10. **Asteroseismology targets** — A follow-up will be performed of all targets in the survey footprint with asteroseismic information from the CoRoT, K2, TESS and PLATO missions, allowing a much more precise determination of elemental abundances, stellar parameters, distances, and ages for these stars (for example, Valentini et al., 2018). These stars are, therefore, also key targets for survey calibration. With the exception of TESS, objects will be distributed in specific fields (from 2.5 square degrees for CoRoT, to 100 square degrees for K2, and 2250 square degrees for PLATO). The brightest of these targets will function as a bad weather programme.

11. **Hot subdwarfs** — The goal of this sub-survey is to compile, classify and analyse the first unbiased all-sky sample of hot subdwarf stars pre-selected on the basis of data from Gaia and photometric surveys from the ground (Geier et al., 2018). Hot subdwarfs are the helium-burning stripped cores of red giants and are likely formed via diverse close binary evolution channels such as stable mass transfer, common envelope ejection and mergers. They are key objects to study interactions between low-mass stars as well as stars and sub-stellar objects. Hot subdwarf binaries qualify as progenitors of type Ia supernovae and calibration sources for gravitational wave detectors. They are chemically peculiar and several classes of pulsators have been discovered which are well suited for asteroseismic analyses. This survey will allow us to perform the first population study of these objects.

### Science requirements

By using the large wavelength coverage and resolution of the LRS, this survey will deliver not only radial velocities with the high precision of 1–2 km s$^{-1}$, but also chemical information, providing individual elemental abundances for iron-peak, α- and n-capture elements, C, Na, and Li (which is detectable in LR only in giants). Depending on the element, elemental abundances will have a precision of 0.1–0.2 dex when sufficient S/N in the continuum is reached in the relevant wavelength regime.

### Target selection and survey area

A careful target selection with a selection function that can be well modelled is crucial for our survey. A large down-sampling is necessary to select ~ 15 million targets from over 300 million Gaia targets in the magnitude range we will explore in 4MIDABLE-LR. Should targets just be picked randomly, we would predominantly select nearby disc stars, which would greatly hamper the effectiveness of our survey to research the Milky Way as a galaxy.

To ensure a homogeneous volume distribution for the main disk coverage, for the sub-catalogues defined in the ESN, Dyn, Chem, and BIG sub-surveys above we use the parallax ($p$) measured by Gaia and its uncertainty in a probabilistic approach, where stars are selected randomly in a given HEALPix[2,b] area such that the probabilistic line-of-sight parallax distribution has the shape of $1/p^4$, implying a distance distribution of the form $d^2$. The ESN is constrained to a cylinder of radius $d$ = 2 kpc and height $|z|$ = 2 kpc, centred on the Sun. The Dyn and Chem sub-catalogues, which share the same volume, are selected from a cylinder with base 2 < $d$ < 15 kpc and height $|z|$ = 2 kpc. The resulting hole around the Sun is complemented by the ESN catalogue. The BIG catalogue covers the volume defined by $|l|$ < 30 degrees and $|b|$ < 20 degrees, and 4 < $d$ < 15 kpc. Given the uncertainties in the measurements of parallaxes, leakage of targets outside these boundaries is both expected and desired.

Our target selection thus balances the need to sample the full Galactic disc, with an attempt to be as unbiased and efficient as possible. Therefore, for the main sub-surveys we foresee a selection based purely on apparent magnitude, parallax, and parallax uncertainties with no additional colour cuts. For some of the smaller sub-surveys there might be colour cuts as they target specific stellar populations. The faint dynamical disc includes a colour selection to target red giant branch stars similar to the selection performed in the 4MOST Consortium Survey Milky Way Halo Low-Resolution Survey.

We intend to use the latest Data Release from the Gaia mission at the time of final 4MOST catalogue submission; the preliminary catalogue that we present here is based on the currently available DR2 (Gaia Collaboration, Brown et al., 2018). An overview of the magnitude range per sub-survey is given in the second column of Table 1.

Unless otherwise mentioned in the first column of Table 1, all sub-surveys cover an area that is all-sky as observed by 4MOST, restricted in declination from –80 < dec < 20 degrees. The areas of dec > 5 degrees and dec < –80 degrees are outside the fiducial survey area and may not get completed (see Guiglion et al., p. 17). As expected, the density of targets is larger along the disc plane (see Figure 1). Areas with extinction so large that the target density is less than the available LRS 4MOST fibres will be avoided. Figures 2 and 3 illustrate that, despite being based on a relatively simple selection of apparent G-magnitude and parallax only (including uncertainties and flags), the resulting distribution of targets in the x-y-z plane in the main sub-surveys is balanced and sufficiently smooth. In Figure 3, we perform the test of resulting homogeneous disc coverage of targets in our Dyn, Chem, and BIG input catalogues using distances obtained with the Bayesian Star Horse code (Queiroz et al., 2018; Anders et al., in preparation), as the targets are too far out to rely on $d = 1/p$ only to calculate distances. However, we stress that these Star Horse distance values are not used in the selection itself.

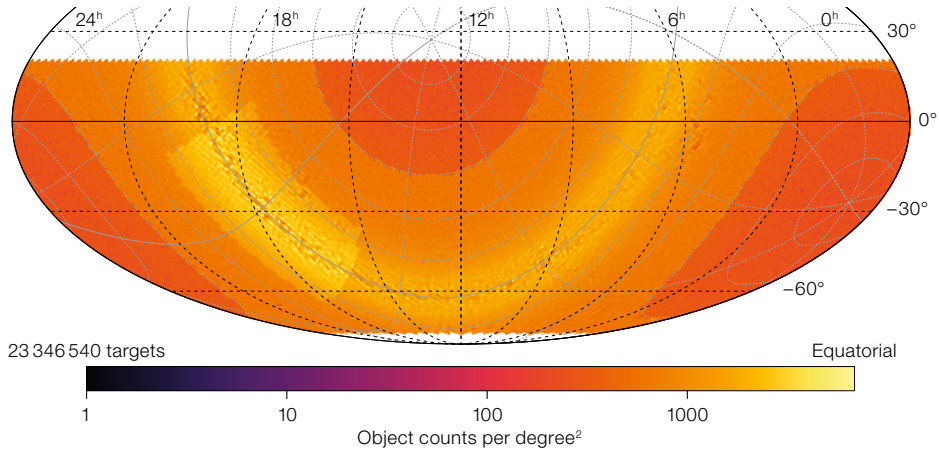

Figure 1. Illustration of the target density of the main sub-surveys in equatorial coordinates. All current catalogues and figures might be subject to change with the further development of our target selection criteria, a better understanding of the interplay of all surveys together, and the implementation of later Gaia data releases. Areas with high extinction will be de-prioritised because the target density drops below the 4MOST fibre density and only nearby targets are observable in these regions. Note that targets with dec > +5 and dec < −70 degrees have only a smaller likelihood of being observed in the fiducial survey strategy (see Guiglion et al., p. 17).

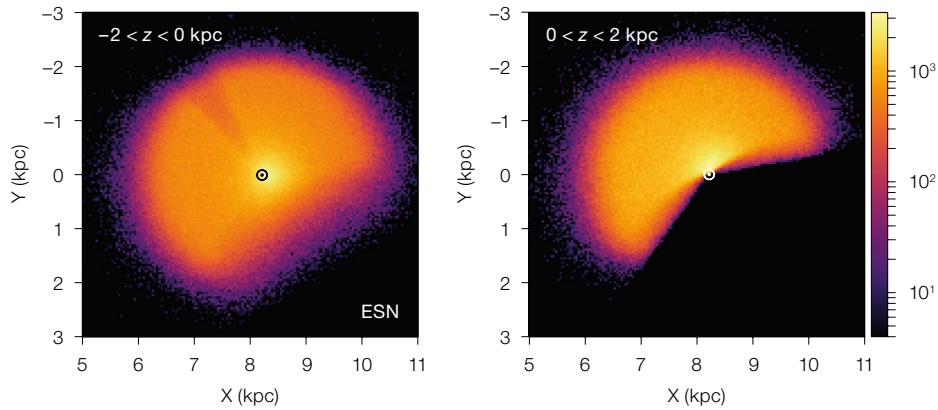

Figure 2. Target density distributions of the current definition of the Extended Solar Neighbourhood (ESN) catalogue using distance $d = 1/p$ for stars below (left) and above the disc mid-plane (right). The selection ensures an almost homogeneous distribution within a cylinder of radius $d = 2$ kpc and height $|z| = 2$ kpc, centered on the Sun. The Galactic centre is at $(x,y) = (0,0)$. The asymmetry between the two plots is due to the restriction in dec of the region of the sky that is observable with 4MOST.

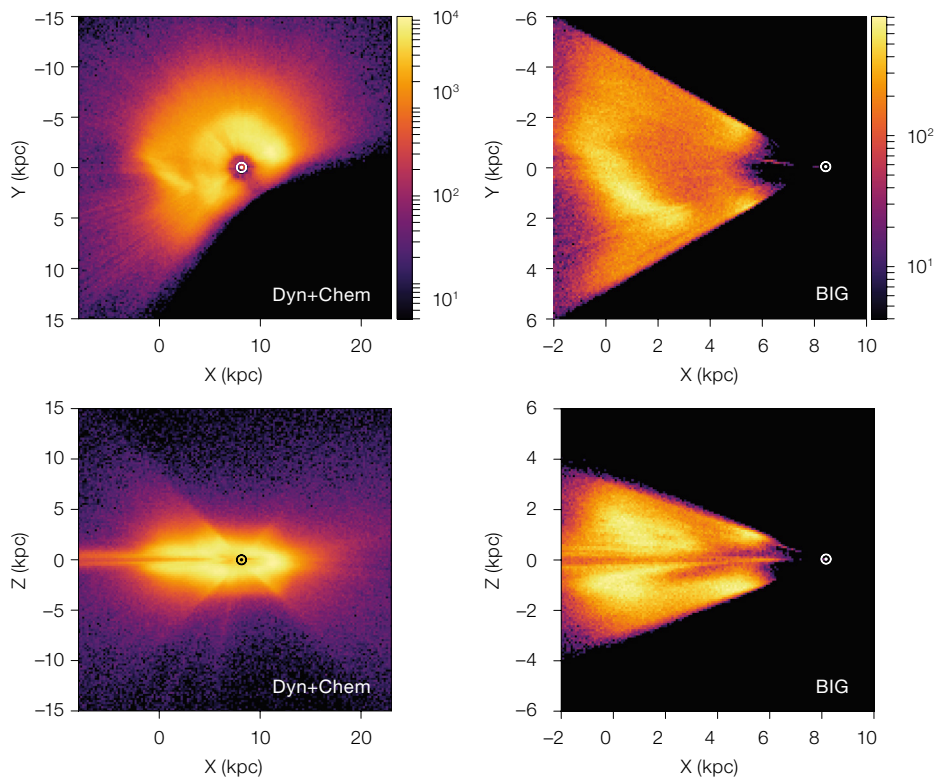

Figure 3. Left column: Face-on (top) and edge-on (bottom) target density distributions of the current disc catalogues. The dynamical (Dyn) and chemodynamical (Chem) disc surveys cover the same area and magnitude range but have different requirements for SNR (see Table 1). The Galactic centre is at $(x,y) = (0,0)$. The artefact at $|b| = 45$ degrees is due to our selection of a cylinder with base $2 < d < 15$ kpc and height $z = 2$ kpc. The resulting hole around the Sun is complemented by the ESN catalogue. Right column: Same as left, but for the Bulge/Inner Galaxy (BIG) catalogue, where our selection is in the volume defined by $|l| < 30$ degrees, $|b| < 20$ degrees, $4 < d < 15$ kpc. For both catalogues we take parallax uncertainties into account and require a homogeneous volume density. Star Horse distance estimates (Anders et al., in preparation; Queiroz et al. 2018) are used for calculating $x$, $y$, and $z$ coordinates. The BIG catalogue clearly shows the presence of the bar (Anders et al., in preparation; Queiroz et al. in preparation).

| Sub-survey name | Gaia (G magnitude) Interval | Spectral success criteria Minimum S/N Å⁻¹ at wavelength interval | | | $N_{min}$ FoM = 0.5 | $N_{goal}$ FoM = 1.0 |
|---|---|---|---|---|---|---|
| Extended Solar Neighbourhood | 14–16.5 | 40 at blue | 50 at MgCa | 20 at Li | $2.5 \times 10^6$ | $4.5 \times 10^6$ |
| Dynamical Disc $|b| < 30°$ | 14–18 | 12 at CaT | 15 at Mgb | – | $3.5 \times 10^6$ | $4.5 \times 10^6$ |
| Faint Dynamical Disc $|b| < 15°$ | 18–19 | 12 at CaT | 15 at Mgb | – | $1.5 \times 10^5$ | $2.8 \times 10^5$ |
| Chemodynamical Disc $|b| < 30°$ | 14–18 | 40 at blue | 50 at Mgb | 20 at Li | $1.5 \times 10^6$ | $2.5 \times 10^6$ |
| Inner Galaxy $|l| < 30°, |b| < 20°$ | 16–17 | 40 at blue | 50 at Mgb | 20 at Li | $8 \times 10^5$ | $1 \times 10^6$ |
| Southern Bulge Deep $|l| < 8°, -10° < |b| < -4°$ | 17–18.5 | 40 at blue | 50 at Mgb | – | $1.2 \times 10^5$ | $1.4 \times 10^5$ |
| Very metal-poor stars | 14–18.5 | 40 at blue | 50 at Mgb | 20 at Li | $1.3 \times 10^5$ | $2.0 \times 10^5$ |
| White Dwarfs | 14–20 | 30 at blue | 30 at Mgb | 30 at CaT | $1.7 \times 10^5$ | $2.0 \times 10^5$ |
| Compact X-ray Binaries | 16–22 | 5 at blue | 50 at MgCa | 5 at CaT | $7.5 \times 10^3$ | $1.5 \times 10^4$ |
| Cepheids | 16–20.5 | 35 at MgCa (G < 18) | – | 10 at CaT | $1.7 \times 10^3$ | $2.3 \times 10^3$ |
| Asteroseismic targets | 4–16 | 40 at blue | 50 at Mgb | – | – | $2.0 \times 10^5$ |
| Hot subdwarfs | 8–19 | 30 at blue | 30 at Mgb | 30 at CaT | – | $2.5 \times 10^4$ |

Table 1. This table provides key information for each of the sub-surveys. Although the information given here reflects the philosophy of the sub-surveys, the exact numbers are subject to change as the target selection advances further in preparation for 4MOST operation. All sub-surveys target the full sky observed by 4MOST, unless specific limits are given in the first column in Galactic longitude (*l*) and latitude (*b*). The second column lists the magnitude range for targets in Gaia *G*-band. The next three columns show the requested minimum S/N per Å for a successfully observed target and the wavelength region(s) — up to three — where this is measured (abbreviated; see text for details). The final two columns list the minimum number of targets successfully observed to reach a FoM of 0.5 for the sub-survey, and the goal number of targets, defining a FoM of 1.0.

## Spectral success criteria

The spectral success criteria are defined by the median S/N ratio per Å in the continuum over a wavelength interval. For the different sub-surveys, different wavelength regions are used for this calculation, depending on the spectral features that are important for the science case. The spectral criteria per sub-survey are given in Table 1; a maximum of three spectral success criteria can be used per sub-survey, as given in the subsequent columns. The wavelength regions are as follows:
– Blue: 4500–4700 Å;
– Mgb: 5140–5200 Å;
– MgCa: 5140–6450 Å;
– Li: 6670–6740 Å;
– CaT: 8350–8850 Å.

The figure of merit (FoM) for each sub-survey is described by the ratio of the number of targets successfully observed ($N_{obs}$) relative to the minimum number of targets and the goal ($N_{min}$ and $N_{goal}$ respectively; see Table 1). Up to $N_{min}$ the relation is linear such that it reaches a figure of merit of 0.5 when $N_{min}$ targets are successfully observed. Thereafter it follows a square root function until it reaches a value of 1 when $N_{obs}$ equals $N_{goal}$. The input catalogues all contain a larger density of targets than $N_{goal}$, to provide operational flexibility. We additionally require a relatively homogeneous coverage of the footprint for our sub-surveys, not allowing for any holes in the footprint that exceed a few square degrees.

Where possible, we favour an observation strategy that allows each field to be observed at least twice with at least a year in between the two observations. This will provide additional information on which stars are in binary systems with detectable radial velocity variability on these timescales. However, in fields with many faint Cepheid targets this strategy is not favourable, since these variable stars will need to be observed consecutively.

### Acknowledgements

Cristina Chiappini acknowledges support from DFG Grant CH1188/2-1 and from ChETEC COST Action (CA16117), supported by COST (European Cooperation in Science and Technology). Else Starkenburg gratefully acknowledges funding by the Emmy Noether program from the Deutsche Forschungsgemeinschaft (DFG). Bertrand Lemasle acknowledges support from the Sonderforschungsbereich SFB 881 "The Milky Way System" (sub-projects A5) of the German Research Foundation (DFG). Sub-survey #7 is funded under the European Union's Horizon 2020 research and innovation programme no. 677706 (WD3D).

### Links

[1] ESA-Gaia mission: http://sci.esa.int/gaia/
[2] HEALPix: https://healpix.jpl.nasa.gov/

### Notes

[a] $R > 5000$ (typically $R \sim 6500$; see de Jong et al., p. 3) is mid-resolution, giving more than twice the spectral resolution of other large low-resolution spectroscopic surveys, for example, the Sloan Digital Sky Survey extension for Galactic Understanding and Exploration (SEGUE) and the Large Area Multi-Object fibre Spectroscopic Telescope (LAMOST).
[b] HEALPix is an acronym for Hierarchical Equal Area isoLatitude Pixelization of a sphere, a pixelisation procedure that produces a subdivision of a spherical surface in which each pixel covers the same surface area as every other pixel.